**Andy Crabtree**
http://orcid.org/0000-0001-5553-6767

School of Computer Science
University of Nottingham
Email andy.crabtree@nottingham.ac.uk

**29 April 2026**

## Author's response to commentaries on H is for Human

This is the author's response to commentaries on the original article *H is for Human and How (Not) to Evaluate Qualitative Research in HCI*, https://doi.org/10.1080/07370024.2025.2475743

Commentaries were provided by:

- Jeffrey Bardzell, https://doi.org/10.1080/07370024.2025.2612474
- Alan Blackwell, https://doi.org/10.1080/07370024.2025.2591878
- Paul Dourish, https://doi.org/10.1080/07370024.2025.2594529
- Bonnie Nardi, https://doi.org/10.1080/07370024.2025.2596752
- Peter Pirolli, https://doi.org/10.1080/07370024.2025.2596745
- Jennifer Rode, https://doi.org/10.1080/07370024.2025.2598800
- Peter Tolmie, https://doi.org/10.1080/07370024.2025.2591872

### Introduction

I would first like to say thank you to the various commentators who have taken time to read, reflect upon, and respond to my article *H is for Human* (Crabtree, 2025), also to the editor for selecting my work for broader consideration, it's an uncomfortable honour I didn't anticipate when I wrote the piece. I hope no offence is taken if I cast the same critical eye over the arguments and advice offered up in the commentaries in response.

Let me begin with Pirolli(2026), who says in light of my arguments that we "should not give up on standards of science." I could not agree more. But what Pirolli fails to grasp is that we need the **right** standards in the first place. That is what *H is for Human* is about. Applying the right scientific standards to qualitative research. It's not an argument about qualitative research versus science. It's an argument about specious versions of science – scientism – built on specious reasoning, and positivism in particular, that puts emphasis on measurement and quantification irrespective of whether or not measurement and quantification are **appropriate** to the evaluation of qualitative research. Having been at the receiving end of many positivistic reviews my feeling was, and still is, that there is need to correct the mundane use of metrics and measurement amongst HCI reviewers in their assessment of qualitative research. But what do the commentaries make of it?

### The methodological minefield

Bardzell (2026) suggests the situation is unproblematic. That besides myself, Lucy Suchman, Terry Winograd, Liam Bannon, Susanne Bødker, and Paul Dourish, to name but a few qualitative researchers active in HCI, have "somehow all managed to get tenure." Bardzell has had "dozens" of qualitative papers published in HCI. He does not believe that qualitative research is "somehow being suppressed." How to respond? I say 'yes', he says 'no' simply leaves us at an impasse. So let me begin with Dourish, who speaks about Dave Randall in his commentary (Dourish, 2026), tells us an anecdote recounted by him to do with ethnography. I'll offer a more recent one: Dave Randall, Mark Rouncefield, Pete Tolmie and I put on an ethnography summer school last year. Dave told the participants about *Faltering from Ethnography to Design* (Hughes et al., 1992), how it had just been given a Lasting Impact Award by the ACM recognising the influence it has had on the field since its publication in 1992. He observed that if you read this paper, **you will not find a methodology section**.

As Randall points out, methodology sections weren't necessarily required when we started out, and if they were, they were often short, practical, and to the point: we went there, we did that, we found this, that sort of thing. They were rarely contentious. We weren't made to jump through hoops. Over the years, however, increasingly detailed methodology sections have become *de rigueur* and with them has come positivism; mundane positivism as Peter Tolmie points out that labours under the yoke of commonsense reasoning and asks how many, how much, how long, etc. We've been encountering and fighting it for years. *How Many Bloody Examples Do You Want?* (Crabtree et al., 2013) was published in response to the application of specious positivist reasoning to ethnographic work and the insistence that observations are only valid if adequately substantiated by measurements (sample size, duration, frequency, etc.). My colleagues and I have run into it many times since, most recently when we submitted *AI and the Iterable Epistopics of Ris*k (Crabtree et al., 2024) to a leading HCI journal which was rejected on the basis that the validity of its "self-evident" findings was "compromised" by a small set of interviews from a small sample; the trigger, as reported, for *H is for Human*.

I was tired of making the same old arguments to reviewers and so here we are having a much grander debate, a "necessary conversation about the philosophy of HCI's practice and its epistemological commitments" as Dourish (2026) puts it, but I don't simply want to echo what I've previously said. To answer Bardzell directly, as Dave Randall makes plain to see, times have changed since the Old School he lists secured tenure. Methodology has effectively been **weaponised**, and not only by mundane positivists but by qualitative researchers too, particularly under the auspices of *reflexivity*. Reflexivity is considered by many in the Human Sciences to be the hallmark of qualitative research in a post-modern world. Nardi (2025) and Dourish (2026) are scornful of what often passes for reflexivity in HCI, dismissing it as an "insult to English grammar" that reduces the relational production of knowledge to a "superficial practice" and "tick box" exercise that merely entails furnishing "positionality statements."

Once again, I couldn't agree more. If one reads my account of reflexivity in *H is for Human* carefully it will be found that I do not argue for positionality statements. Rather, I emphasize the provision of a *practically adequate* methodological account that enables reviewers to understand how the researcher(s) interacted with their participants to produce the knowledge (findings, results, insights) described in a paper. This is not a call for an 'an account about me' to paraphrase Dourish (ibid.), but for a description of the **practicalities** of engaging participants and arriving at some understanding of their world. It is not an invitation to reflect on the self and retreat to theoretical considerations, but to consider one's actual relationship to one's participants and **how** they were practically engaged in the production of knowledge. Rode (2026) would have me go much further. Reflexivity for Rode is a general phenomenon that cuts through fieldwork, analysis, and the production of the text. As we have previously elaborated at length in *Deconstructing Ethnography* (Button et al., 2015), writing accounts 'about me' – how I relate to and interacted with participants, how I gathered data, how I analysed it, how I found meaning in the data, how I wrote up my findings, etc. – does not mean that I necessarily understand the people I have been studying; indeed reflexive accounts 'about me' might very well (and all too aften do) **get in the way** of apprehending 'the other'.

Rode's commentary offers an anthropological account of ethnography and the role of reflexivity in a discipline that attempts to study people that are in significant ways *not* like ourselves. It overlooks the much more influential and far-reaching use of ethnography in sociology championed by the Chicago School. The Chicago School was an intellectual powerhouse of 20th Century social thought and inquiry. It drove the development of qualitative research methods in the 1920s (Platt, 1985), including ethnography, and Chicago-style ethnographies continue to flourish (Deegan, 2001). Ethnography may have originated in colonial studies of tribal peoples (Hammersley, 2017) but its use in sociology to understand people in our own societies and culture *outstrips* anthropology by several orders of magnitude, if for no other very practical reason than few people are actually prepared to dedicate years of their life living with the natives in faraway places, in contrast to spending considerably shorter and more comfortable periods of time hanging out in their own backyard. The kind of ethnography championed by Hughes et al. (1992) in *Faltering from Ethnography to Design* owes far more to the Chicago School than it does to anthropology. If Dourish is serious about HCI being a "sociological enterprise" (Dourish, 2026), and I sincerely hope he is and that he will in due course expand on the point, then HCI researchers will need to adopt a much more critical stance towards anthropological configurations of ethnography.

The problems encountered in engaging with people who really are different to us – who speak a different language and live a very different even alien way of life – are very different to and require different solutions to the problems encountered in using ethnography to study people in our own societies, i.e., **'members'** (Garfinkel and Sacks, 1969). It's not clear how anthropological constructs of reflexivity that have emerged from efforts by *non-members* to study *other* peoples are generally appropriate to the sociological task and empirical elaboration of membership competence in all manner of circumstances and domains, or that they should be uncritically accepted as they largely have been in HCI to date. Furthermore, Rode's commentary leaves untouched many other kinds of reflexive construct. As I said in *H is for Human*, Lynch (2000) identifies over *twenty* different kinds of reflexivity in the Human Sciences more generally, and there is no reason to expect that number to have decreased in the last quarter of a century. Doing qualitative research in HCI is becoming a **methodological minefield**. It is very much potluck as to what kind of reviewers one may get, including ones wedded to positivism, or reflexivity, or even to both, and what methodological demands they will make.

## Against method

I am reminded of a paper written by the late Egon Bittner (1973), where he reflects on fieldwork and the inadequacies of objective and realist stances alike. Bittner carefully lays out how "for the field worker" – always a "visitor" in someone else's world – "things are never naturally themselves but only *specimens* of themselves", "*exhibits*", subsequently drawn upon to render "someone's social reality", typically and ironically in ways that "far from being realistic are actually heavily intellectualised constructions." Bittner shows us that neither objective nor realist stances offer a procedural remedy to the fieldworker's "distortive tendencies", which as Garfinkel (1996) elaborates, put emphasis on "constructive analysis" (e.g., modelling) and "generic theorizing." Similarly, neither positivism *or* reflexivity offer a solution either, and not only to fieldwork but to any kind of qualitative research, because **no procedural formula exists** in the Human Sciences, no scientific method, for rendering someone else's social reality in ways that retain the sense of "things as they actually present themselves to the perceiving subject", i.e., as they naturally occur, are encountered, and experienced (Bittner, 1973).

Like Bittner, and as Shakespeare puts it for that matter, I would cast **a plague o' both your houses**, as neither positivism nor reflexivity get us any closer to solving the problem of how to "do justice to the realities" we study (Bittner, 1973), to someone else's world, or at least those parts of their world we have been fortunate enough to be let into and allowed to study, whether through fieldwork, interviews, or some other means; constructive analytic and theoretical tendencies continually render social reality **elusive** (Pollner, 1987). However, there is a quote in a paper by Mike Lynch where he's discussing ethnomethodology's rejection of theorising as an essential mode of interpretation (contrary to Rode's insistence) that I find instructive:

> David Sudnow asserts that an ethnomethodological "finding" has to do with "the nature of the world" and not "the nature of the procedure" through which the world is observed. He goes on to say, "if the world, in fact, is general, it is general whether you get it by induction or however else you might get it. It is the world that provides the adequacy for the generalizations you can make about it and nothing else." (Lynch, 1999)

I take Sudnow and Lynch's methodological reflections seriously: **however** you find it will do. I would argue that we need to **stop regulating how** to find the world and understand other people's social realities, stop trying to impose inappropriate methodological procedures based on mythologies of science: e.g., that there is a scientific method, that science progresses through doubt, that replication is a hallmark of science, etc. (Lynch, 1993). Even if the practices of Natural Science were understood – and they aren't as yet (see Garfinkel, 2022) – they would *not be relevant* to Human Science as I explained in *H is for Human*. Reflexivity "offers no guarantees" either as Lynch (2000) makes clear. What chance would a paper like *Faltering from Ethnography to Design* have in today's methodological minefield? We should **dispense with methodological stipulations**. Stop making people jump through a random array of hoops assembled by whatever random array of reviewers get randomly assigned to a paper, and leave it instead to researchers of whatever hue to tells us what worked for them. Insofar as qualitative research is concerned, we can determine *if it works for us* by applying the evaluation criteria elaborated in *H is for Human* (see below).

The suggestion that we should abandon methodological prescriptions is not mine but should rightfully be recognised as philosopher of science, Paul Feyerabend's. Feyerabend opens his (in)famous essay *Against Method* in saying, "anarchism, while perhaps not the most attractive political philosophy, is certainly excellent medicine for epistemology, and for the philosophy of science (Feyerabend, 1975)." HCI might benefit from a spoonful or two right now. Feyerabend's "anarchistic methodology" only makes sense against a background of methodological quarrels and claims that there is *a* correct method "for conducting the business of science", whether its positivism or reflexivity or something else. It is against this background that Feyerabend says "anything goes." Feyerabend's argument is not that that we should dispense with science, but that we should dispense with formulaic accounts of science, as science doesn't actually work this way, it is much more anarchic in practice. It would be more appropriate, then, to focus our efforts on the business rather than the philosophy of science and **rigour not formula**.

### H is for Human Science

Rigour, as Blackwell (2025) hints at, is enabled through a heterogenous array of research practices in the Human Sciences. It is surprising then that he seems somewhat taken aback by the idea that one would invoke such a thing as Human Science as somehow relevant to HCI, especially since Blackwell championed the notion of HCI as an inter-discipline (Blackwell, 2015). Indeed, he has previously argued that HCI is "fundamentally interdisciplinary", "defined, not as a subject in itself", and "not a scientific discipline", but a "professional endeavour" defined "in relation to other disciplines" (ibid.). It might not be unreasonable to expect Blackwell would know something about what HCI sits *in-between* then. Apparently not. Indeed, Blackwell has never heard of the term Human Science, an "obscure (100 year-old) brand" that he had to look up, unworthy of pedagogical discussion as "we live in a world where new ideas come from making" that appears to have started sometime in 1997. Blackwell goes on to argue that "qualitative and interpretive methods are essential" to HCI; not only "methods of practice-led and artistic inquiry", but also "the rigorous social sciences of anthropology, sociology, education, public policy, economics and others." The **Human Sciences**, in other words.

The Human Sciences are not all of apiece. They do **not** all adhere to the **same** standards of rigour. Nardi (2025) suggests that ethnography, "a core qualitative methodology", is "not averse to including quantitative analysis under its interpretive umbrella." But ethnography is not *a* core qualitative methodology. As Hammersley (2017) reminds us,

> "Ethnography is not a single method or a single approach: it is perhaps best thought of as a *family* of approaches**.** While some of the differences reflect variations in the phenomena being investigated, they also stem from fundamental disagreements about the nature of the social world, how it can be understood, and what form social research should take."

It would be better, indeed more accurate, to say that some ethnographers adopt a mixed methods approach, than to make claims for ethnography *per se*. As I said in *H is for Human*, we need to be very careful about such approaches. We need to be sure that we respect the **fundamental differences** between the quantitative and the qualitative. Otherwise the blurring of methods allows inappropriate reasoning to seep into and colour the evaluation of qualitative research.

Thankfully none of the commentators wanted to defend the idea that **reproducibility** is an appropriate scientific standard. Let the bells ring.

Nardi thinks there nothing wrong with **generalisation**. Neither do I. I only dispute the way it is construed of positivistically, as Nardi does, as something that *requires* a "larger, well-drawn sample" in contrast to something that is built sociologically into even a **single** case, as elaborated in *H is for Human* and as my colleagues and I have demonstrated empirically elsewhere (Crabtree et al., 2013).

Nardi and Pirolli take issue with idea that the insights produced through qualitative research should be 'apodictic', i.e., self-evident, **given in the empirical material provided**, and thus available for anyone including reviewers to see. For Nardi, apodicity means "there would be no disagreements in the literature." I think the heterogenous character of Human Science means the literature is and always will be diverse. But what the literature has to do with evaluating the specific results or claims of a qualitative study, and whether or not they are clearly supported by the empirical material provided, is glossed over. Whether or not Nardi thinks results or claims should be visibly evidenced and **plain to see** in qualitative research is passed by. It's hard to see how and why anyone would want to argue against apodicity.

Nonetheless, Pirolli tries in saying that "science does not progress by being apodictic, but by being strenuously doubted and not being wrong." The idea that doubt is essential to science is not novel, even Alred Schutz mythologised science (Lynch, 1988). However, Edmund Husserl, who exerted considerable influence on the interpretive tradition independently of Schutz, tells us that the natural scientist labours under the same fundamental assumption as the layman, i.e., both take the world and the possibility of knowledge of the world as given. Fundamentally, there is no possibility of doubt for either party (Husserl, 1999). The scientist, like the layman, may **occasionally** doubt their observations and findings, but doubt cannot be a primary driver of scientific knowledge, infinite dubitable regress would be fatal to the enterprise. There has to be 'bedrock' somewhere (Wittgenstein, 1975).

Pirolli's comments may best seen, then, as a riposte to my instruction that reviewers should not try to conceive of ways to doubt the results of a qualitative study or imagine how they could be different **if** they find them self-evident, because (to reiterate) "science progresses by scepticism." However, just as with reproducibility, it would be better and indeed more apposite to treat scepticism as something that has to be warranted by circumstance, not done for the sake of it. Just because one *can* doubt something – any philosopher worth their salt will argue that black is white and white black, for example – doesn't mean we should *try* to doubt things as a matter of course. As Newstead puts it, to do so would be to act in **"bad faith"** (Newstead, 2026).

Apodicity is not naughty advice, reviewers *can* believe their eyes, they should not appeal to extraneous criteria and measurements instead. What they should do is ask themselves if the results or claims of a qualitative study are **visibly** supported by the empirical material provided If they are, good. If they aren't – and believe me, I have been reviewing qualitative research inside and outside HCI for many years and I can testify that a great many qualitative studies do not offer apodictic insight – then it is not unreasonable to reject them. Just don't act in bad faith.

### How to evaluate qualitative research

I'm sure those who apply quantitative measures to qualitative research would say they are being rigorous but that's a very limited and inappropriate view on what constitutes rigour. As previous efforts to define a rigorous evaluation framework for qualitative research show (Spencer et al., 2003), it will have to deal with enormous variety and complexity, which leads us to consider what an appropriate treatment of qualitative research looks like? For Bardzell (2026), at "logical extreme" a paper like *H is for Human* would be required to account for "ever more niche methodologies and practices", unless a "stopping rule" can be found. I don't agree. I offered 5 criteria for reviewers to use to determine whether or not a qualitative study **in general is acceptable** to HCI regardless of niche methodologies or practices.

- Can the reviewer understand the methodological approach employed, how the researcher(s) practically engaged with participants so as to produce the knowledge (findings, results, insights) described in a study?
- Are the findings apodictic, are they given (self-evident) and plain to see in the empirical material provided?
- Does the paper elaborate, extend or refine an existing sensitizing concept or introduce a new one that furnishes novel analytic insight?
- Does the analytic insight provided have reach or potential utility and can it thus be used to develop our understanding of some aspect of human-computer interaction in today's world?
- Is the paper of relevance to HCI, has the case actually been made?

Together these criteria constitute a minimum standard of rigour for evaluating qualitative research that dispenses with extraneous measures and is agnostic to an ever-growing arsenal of reflexive constructs in the Human

Sciences. I don't think it especially stringent. Rode (2026) would clearly prefer I go further, for example, and I'm sure others will too. **Evidently, there is no stopping rule**.

Bardzell is asking the wrong question. The question isn't about where it stops, but how it starts? How are HCI reviewers, seasoned veterans and novices alike, to arrive at a point where they **acknowledge they might be the problem that needs fixing**? Peter Tolmie (2025) is quite right when he says that many reviewers "may not believe they have a positivist orientation because they have not had a formal positivist training ... how, then, does one make them stop a moment and say, 'hang on a minute'?" The solution is not evident to Tolmie, nor to me, though he suspects it may require "wholesale revision of the current apparatus for peer review." I don't know if *H is for Human* represents such a radical move. I do know that it is necessary to **promote awareness** of the inappropriate methodological treatment of qualitative research in HCI as a necessary precursor to change. Writing articles and commentaries about it are one mechanism for doing this, but obviously not sufficient in themselves. Other forms of **concerted action** will also be required. It might be that publishers and professional associations come to play a formative role, but that will require significant push by senior leaders in the field and undoubtedly take time. More tangibly, journal editors and associate editors, conference organisers and theme chairs, could play a significant role and take steps to issue appropriate guidance to reviewers (a short bullet-list is provided above) and ensure compliance, but I wouldn't hold my breath waiting for that to happen consistently and at scale.

The individual also has a role to play. Indeed, **the author might be of immediate influence** in highlighting for reviewers what an appropriate treatment of their work should look like. Why not write it into your research? Complement your positionality statement (if you feel you must provide one) with a statement describing **How To Assess This Research**. Make reviewers aware that there is a common problem in the treatment of qualitative work on HCI. A problem that is beginning to be recognised in HCI. A problem that reviewers need to be aware of and avoid repeating. Explain the problem in your papers, not in response to reviewer comments. Try and lift the scales from their eyes **before they make a decision**, reviewers really don't like changing their minds, it doesn't look good, too much loss of face involved. Draw on and reference the work of Braun & Clarke (Clarke et al., 2025), Soden et al. (2024), *H is for Human* (Crabtree, 2025) and anyone else who helps you make the case. Make the effort. It might be a while before anybody else acts on your behalf, and it might be that a groundswell is sufficient to drive the change that is required. Good luck.

## Declaration of funding

This work on which this article is based was supported by the Engineering and Physical Sciences Research Council grant number EP/V026607/1.